\documentclass[
reprint,
bibnotes,
amsmath,amssymb,
aps
]{revtex4-2}

\usepackage{graphicx}
\usepackage{dcolumn}
\usepackage{bm}
\usepackage{hyperref}
\usepackage[version=4]{mhchem}    
\usepackage{csquotes}
\hypersetup{colorlinks=true,linkcolor=red,citecolor=red}
\usepackage{float}
\usepackage{xspace}

\usepackage{prettyref}
\newcommand{\pref}[1]{\prettyref{#1}}
\newrefformat{fig}{Fig.~\ref{#1}}
\newrefformat{tab}{Table~\ref{#1}}
\newrefformat{sec}{Sec.~\ref{#1}}
\newrefformat{app}{App.~\ref{#1}}
\newrefformat{eqn}{Eq.~(\ref{#1})}

\usepackage{bbding} 

\usepackage[dvipsnames]{xcolor}

\newcommand{\sco}{SrCrO$_3$\xspace}
\newcommand{\tg}{${t_{2g}}$\xspace}
\newcommand{\dispro}{${2d^3 +d^0}$\xspace}
\newcommand{\dftu}{DFT+\,$\mathcal{U}$\xspace}

\begin{document}

\title{Emergence of a potential charge disproportionated insulating state in \sco}

\author{Alberto Carta}
\author{Anwesha Panda}
\author{Claude Ederer}
\affiliation{Materials Theory, ETH Z\"urich, Wolfgang-Pauli-Strasse 27, 8093 Z\"urich, Switzerland}

\date{\today}

\begin{abstract}
We use a combination of density functional theory (DFT) and dynamical mean-field theory (DMFT) to investigate the potential emergence of a charge-disproportionated insulating phase in \sco, whereby the Cr cations disproportionate according to $3\text{Cr}^{4+} \rightarrow 2\text{Cr}^{3+} + \text{Cr}^{6+}$ and arrange in ordered planes perpendicular to the cubic [111] direction.
We show that the charge disproportionation couples to a structural distortion where the oxygen octahedra around the nominal Cr$^{6+}$ sites contract, while the octahedra surrounding the Cr$^{3+}$ sites expand and distort.
Our results indicate that the charge-disproportionated phase can be stabilized for realistic values of the Hubbard $U$ and Hund's $J$ parameters, but this requires a scaling of the DFT+DMFT double counting correction by at least 35~\%. 
The disproportionated state can also be stabilized in \dftu calculations for a specific magnetic configuration with antiparallel magnetic moments on adjacent Cr$^{3+}$ sites and no magnetic moment on the Cr$^{6+}$ sites.
While this phase is higher in energy than other magnetic states that can arise in \sco, the presence of an energy minimum suggests that the charge disproportionated state represents a metastable state of \sco. 
\end{abstract}

\maketitle
\section{\label{sec:Intro}Introduction}

The electronic properties of many materials with partially filled $d$ or $f$ shells are characterized by strong electronic correlations, which can lead to peculiar phenomena such as, e.g., metal-insulator transitions, superconductivity, or non-Fermi-liquid behavior~\cite{Imada/Fujimori/Tokura:1998, Dagotto:1994, Stewart:2001}.
The basic low-energy physics of these systems can often be understood in terms of an effective multi-orbital Hubbard model, where electronic correlations emerge as a consequence of the local Coulomb repulsion $U$ and the Hund's interaction $J$~\cite{Hubbard:1963, Kanamori1963, Georges2013}.
A strong $U$ generally favors the Mott-insulating state, where the electron number fluctuations on any given site away from the nominal integer filling are suppressed by the strong Coulomb repulsion~\cite{Austin1970, Imada/Fujimori/Tokura:1998, Fazekas:1999}. The Hund's coupling $J$ also plays an important role by favoring a local spin polarization. While, at half-filling, $J$ cooperates with $U$ in the formation of the Mott insulator, away from half-filling, $J$ can compete with $U$ and give rise to strongly correlated metallic phases~\cite{Haule/Kotliar:2009, deMedici2011, Georges2013}.

For strong enough $J$ and not too large $U$, the Hund's coupling can also favor the formation of a charge disproportionated insulating phase, where the initially symmetry-equivalent correlated sites separate into two or more sublattices with different valences~\cite{Strand2014, Subedi2015, Hampel2019, Isidori2020, Merkel/Ederer:2021}. Such charge disproportionation can be observed in several transition metal perovskite oxides, for instance, in rare-earth nickelates~\cite{Alonso1999, Alonso_et_al:2000}, CaFeO$_3$~\cite{Takano_et_al:1977}, and other ferrites~\cite{Battle1990, Guo2017}. The charge disproportionation is usually accompanied by a structural ``breathing'' distortion of the oxygen octahedra surrounding the transition metal cations, where the oxygen octahedra surrounding the more occupied cations expand and those around the less occupied sites contract~\cite{Alonso1999, Alonso_et_al:2000, Woodward_et_al:2000}. 
This strong coupling between the electronic and structural degrees of freedom can play a crucial role in stabilizing the disproportionated state~\cite{Subedi2015, Mercy2017, Hampel/Ederer:2017, Hampel2019, Peil2019, Merkel/Ederer:2021, Georgescu2022}. 

The physics of charge disproportionation in the rare-earth nickelates, and to some extent also in CaFeO$_3$, can be understood in terms of an effective model considering only the $e_g$ orbitals, which disproportionate according to $2 e_g^1 \rightarrow e_g^2 + e_g^0$~\cite{Subedi2015}. While for the nickelates, the lower-lying $t_{2g}$ orbitals are completely filled, they are half-filled in the case of CaFeO$_3$, but can to some approximation be treated as localized $S=3/2$ spins~\cite{Merkel/Ederer:2021}. 
In both cases, the strong hybridization between transition metal $d$ and O $p$ states, and the small or even negative charge transfer energy in materials containing late $3d$ cations with high oxidation states have been proposed as being favorable, or even crucial, for the charge disproportionation~\cite{Khomskii:2014, Johnston_et_al:2014, Rogge_et_al_PRM:2018}.

A recent study of the three-orbital Hubbard model by Isidori \textit{et al.}~\cite{Isidori2020} suggests that in principle similar physics can also emerge in systems with partially filled $t_{2g}$ orbitals. In particular, for an average filling of two electrons per site, a disproportionation according to $3d^2 \rightarrow$ \dispro can occur. 
However, it is unclear whether specific materials in the required parameter regime can be identified. Furthermore, the model studied in  Ref.~\cite{Isidori2020} was based on a featureless density of states without a specific underlying lattice structure. So the question on how the spatial arrangement of inequivalent sites might affect the stability of the corresponding charge disproportionated state remains open. 

The perovskite chromates, $A$CrO$_3$, where $A$ can be Ca, Sr, Ba or Pb, contain nominal Cr$^{4+}$ with two electrons in the \tg manifold. This and the high oxidation state of the Cr makes this family of compounds ideal candidates to explore a potential charge disproportionation in the \tg shell.
In the case of PbCrO$_3$, there have indeed been reports of charge disproportionation accompanied by insulating behavior~\cite{Wu_et_al:2014, Cheng2015, Zhao_et_al:2023}. However, the fact that Pb can also disproportionate into Pb$^{2+}$ and Pb$^{4+}$, where Pb$^{2+}$ hosts a steorechemically active lone-pair, makes this case significantly more complex and results in partially conflicting reports~\cite{Yu_et_al:2015, Paul_et_al:2019}.

For the Ba compound, there is experimental evidence showing insulating behavior in thin films~\cite{Zhu2013_bacro3}, which, however, has been suggested to arise as a consequence of orbital and magnetic order, either in combination with a Jahn-Teller structural distortion or driven by spin-orbit coupling~\cite{Giovannetti2014, Jin2014}.

Partially conflicting experimental reports are available regarding the metallic or insulating character of both the Ca and Sr compounds~\cite{Chamberland1967_first_experiment, Zhou2006_NM_insulator, OrtegaSanMartin2007_partial_OO_experimental, Komarek2011_experimental_c_type, Long_et_al:2011, Cao2015_paramag_semicond}, and a strain-induced metal-insulator transition has been reported for \sco thin films~\cite{Bertino2021_MIT_strain}.
In a previous work~\cite{Carta2022}, we have shown that tensile strain promotes a Jahn-Teller distorted and orbitally-ordered insulating state in \sco, which is compatible with this reported strain-induced metal-insulator transition.

Here we explore whether charge-disproportionation can provide an alternative mechanism for the formation of an insulating state in \sco. We mainly use density functional theory in combination with dynamical mean-field theory (DFT+DMFT)~\cite{Georges_et_al:1996, Kotliar/Vollhardt:2004, Held2007, Lechermann2006, Pavarini_et_al:2011}, which provides a straightforward way to analyze the effect of local Hubbard-type interactions in the paramagnetic state, and also allows for a systematic comparison with the model study from Ref.~\cite{Isidori2020}.
We find that a charge disproportionated insulating state can indeed be stabilized in \sco for realistic interaction parameters, but its stability depends on a subtle balance between the local interaction treated in DMFT and the DFT response to the DMFT charge-density correction, and requires a scaling of the double counting correction for the local interaction included both at the DFT and DMFT level.
Further \dftu calculations suggest that the charge disproportionated state is not the ground state of the system, with different magnetically and orbitally ordered states exhibiting lower energies. 
Nevertheless, charge-disproportionation in \sco could potentially be stabilized by appropriate tuning of suitable parameters such as, e.g., strain, pressure, ionic substitution, etc.

The remainder of this work is organized as follows: in \pref{sec:Methods} we first specify the specific spatial arrangement of inequivalant sites and the resulting unit-cell used to represent the charge disproportionated state, then we describe the corresponding structural breathing mode distortion. After that, we give a brief summary of the DFT+DMFT method and specify all parameters used in our calculations. We then start \pref{sec:Results} by discussing results of simple one-shot DFT+DMFT calculations, followed by an analysis of the significant changes introduced by charge-self consistency. We subsequently explore the interplay of both lattice distortion and the DFT+DMFT double-counting (DC) correction on the stability of the charge disproportionated insulator. Finally, we also present results of \dftu calculations, which allow to perform full structural relaxations and provide alternative insights in the stability of charge disproportionation in \sco.

\section{\label{sec:Methods} Methods}

\subsection{Choice of unit cell and distortion mode}
\label{sec:structure}

\begin{figure}
   \centering
   \includegraphics[width=\columnwidth]{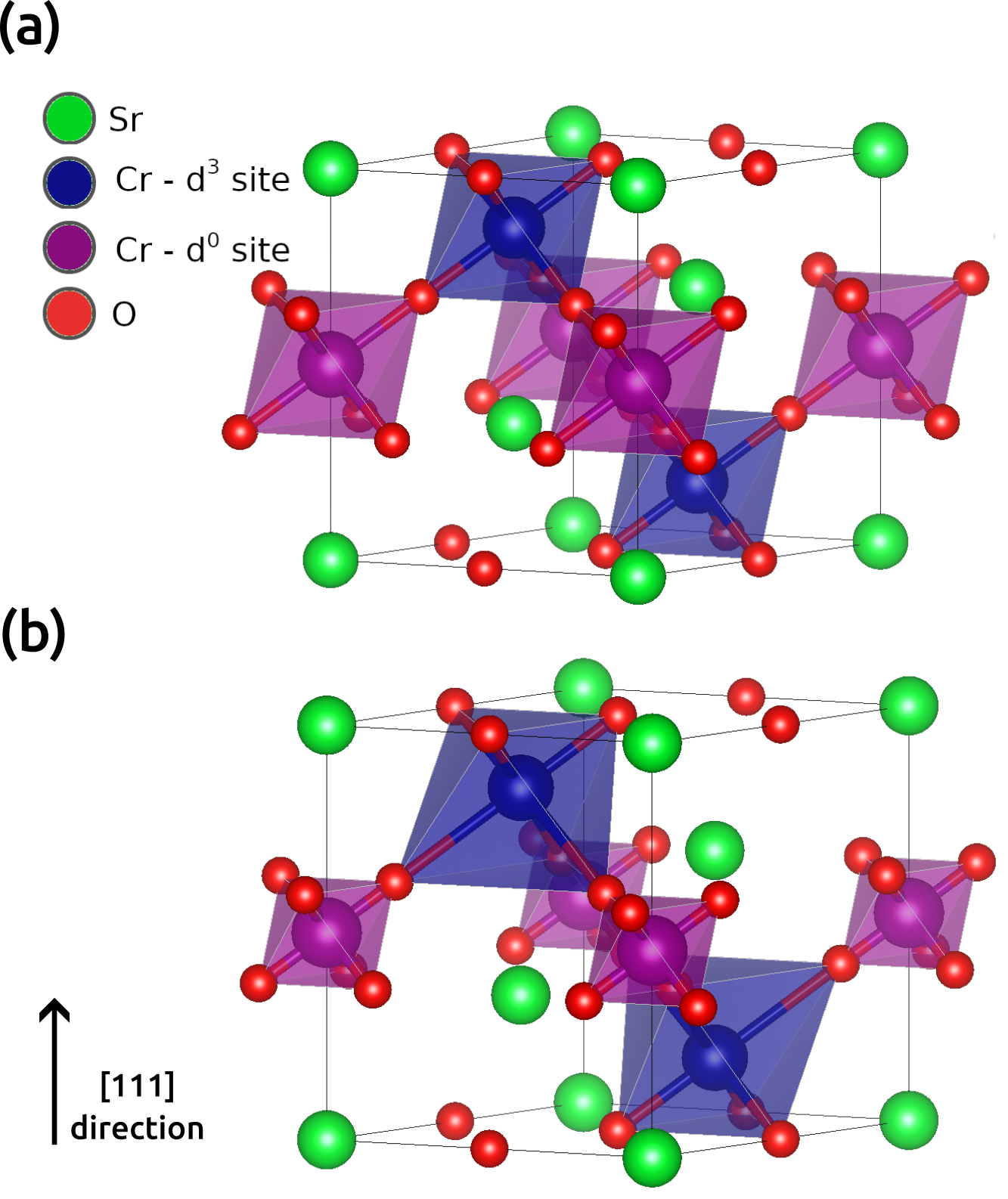}
   \caption{\textbf{(a)} Structural arrangement of inequivalent Cr sites in our unit cell allowing for the emergence of charge disproportionation into nominal $d^3$ and $d^0$ Cr sites in otherwise cubic \sco. \textbf{(b)} Depiction of the $\Lambda_1$ distortion: the octahedra around the nominal $d^0$ sites contract and the octahedra around the $d^3$ sites expand and distort. The thin black lines in both panels indicate the hexagonal unit cell containing three formula units of \sco.}
   \label{fig:3d_dispro_distor}
\end{figure}

To explore the possibility for charge disproportionation in \sco, we need to use a unit cell that can accommodate possible arrangements of inequivalent Cr-sites with a ratio of two nominal Cr$^{3+}$ ($d^3$) sites per one nominal Cr$^{6+}$ ($d^0$) site.
The optimal spatial arrangement of such a disproportionated phase is not clear {\it a priori}.  Here, we use the arrangement shown in \pref{fig:3d_dispro_distor}(a), which consists of two potential Cr $d^{3}$ layers followed by one potential Cr $d^{0}$ layer arranged perpendicular to the cubic $[111]$ direction and repeated periodically. This results in a tripling of the conventional cubic perovskite cell, which can be described in a hexagonal setting with the cubic $[111]$ direction defining the hexagonal axis. This specific arrangement is motivated as the configuration that minimizes the nearest neighbor Coulomb interaction, since each $d^{0}$ site is surrounded by six $d^{3}$ nearest neighbors, whereas each $d^{3}$ site is surrounded by three nearest neighbors of the same kind and three (i.e., the maximum possible amount given the 2:1 ratio) of the other ionic species. 
Furthermore, the same disproportionation pattern has also been observed in La$_{1/3}$Ca$_{2/3}$FeO$_3$~\cite{Guo2017}, where the Fe cations disproportionate according to $3 \mathrm{Fe}^{3.67+} \rightarrow 2\mathrm{Fe}^{3+}+\mathrm{Fe}^{5+}$.
The space group symmetry in the charge ordered phase then reduces from cubic $Pm\bar{3}m$ to rhombohedral $P\bar{3}m1$.

As mentioned in the introductory section, the transition to a charge-disproportionated insulating phase in other perovskite oxides is generally accompanied by a structural ``breathing mode'' distortion consisting of an expansion and contraction of the oxygen octahedra surrounding the cation sites with higher and lower $d$ occupation, respectively \cite{Alonso1999, Alonso_et_al:2000}. Furthermore, it has been shown that the coupling between electronic degrees of freedom and this octahedral breathing mode is crucial for stabilizing the charge disproportionated insulating state in the rare earth nickelates~\cite{Subedi2015, Mercy2017, Hampel/Ederer:2017, Hampel2019, Peil2019, Georgescu2022}.

Since we expect similar physics to be relevant for a potential charge disproportionation in \sco, we also study the coupling to an analogous octahedral breathing mode.
However, in the proposed arrangement, there are always two $[111]$ planes of $d^3$ sites adjacent to each other. Since the corresponding oxygen octahedra are connected, the $d^3$ octahedra cannot expand uniformly without distorting.
As a result, the oxygen octahadra surrounding the nominal $d^3$ sites exhibit three longer and three shorter Cr-O bond lengths, whereas the $d^0$ octahedra can exhibit a simple uniform contraction.

The resulting mode with symmetry label $\Lambda_1$ is shown in \pref{fig:3d_dispro_distor}(b), and can be understood as a simple displacement of the oxygen atoms in between the $d^3$ and $d^0$ sites along the Cr-O bond direction. 
The amplitude of this distortion can thus be parameterized by specifying only one degree of freedom describing the contraction of the octahedra around the $d^0$ sites:
\begin{equation}
     Q = \frac{\ell^0_\text{Cr-O}-\ell_\text{Cr-O}}{\ell^0_\text{Cr-O}} \quad .
    \label{eqn:distortion}
\end{equation}
Here, $\ell_\text{Cr-O}$ and $\ell_\text{Cr-O}^0$ are the Cr-O distances of the $d^0$ octahedra in the distorted and undistorted states, respectively, where the latter is equal to half the Cr-Cr distance $\ell^0_\text{Cr-O} = \ell_\text{Cr-Cr}/2$. 

The $\Lambda_1$ mode leads to a symmetry lowering compatible with the $P\bar{3}m1$ space group. It can thus be assumed that this mode is the energetically dominant structural distortion coupling to the charge disproportionation, and we investigate the effect of the $\Lambda_1$ mode in \pref{sec:results-dmft-distortion} using DFT+DMFT calculations. 
However, the $P\bar{3}m1$ space group symmetry also allows for a number of other distortion modes relative to the ideal cubic perovskite structure, which could in principle lead to a further energy lowering of the charge disproportionated state. 
Therefore, to verify the dominant role of the $\Lambda_1$ distortion mode for the potential charge disproportionation in \sco, we also perform full structural relaxations using the \dftu method~\cite{Anisimov/Zaanen/Andersen:1991,Himmetoglu2013}, which are presented in \pref{sec:dft+u}. 

\dftu calculations have been applied to rare earth nickelates and charge disproportionated ferrites, but these calculations require a local spin splitting in order to provide realistic crystal structures for these systems~\cite{Hampel/Ederer:2017,Zhang_et_al:2018,Dalpian_et_al:2018}.
The simplest way to introduce such a local spin splitting without any further symmetry-lowering is to impose a ferromagnetic (FM) order in the material. However, superexchange between two neighboring $d^3$ Cr sites is expected to lead to an antiferromagnetic coupling, whereas the nominal $d^0$ Cr has no magnetic moment. Therefore, we also explore a peculiar magnetic pattern, where the two nominal $d^3$ Cr sites are initialized with opposite spins and the nominal $d^0$ Cr is initialized without a magnetic moment, leading to a $[ \, \cdots \uparrow 0 \downarrow \uparrow 0 \downarrow \cdots \,]$ arrangement of spins along $[111].$ We refer to this as ``d-AFM'', which stands for \textit{disproportionated antiferromagnetic} phase.

\subsection{DFT+DMFT for \sco}
\label{sec:dmft}

Within DFT+DMFT (and \dftu), the electronic degrees of freedom are separated into the \emph{correlated subspace}, for which the strong local electron-electron interaction is incorporated explicitly, and the rest, which is treated within the usual semi-local approximations of DFT. 
The Hamiltonian then takes the following form~\cite{Held2007, Lechermann2006, Pavarini_et_al:2011}: 
\begin{equation}
\hat{H} = \hat{H}_\text{KS} + \hat{H}_\text{int} - \hat{H}_\text{DC} , 
\label{eqn:full_hamiltonian}
\end{equation}
where $\hat{H}_\text{KS}$ is the effective single particle Kohn-Sham Hamiltonian, which contains the kinetic energy as well as the interaction with and among the uncorrelated electrons and the ionic cores, $\hat{H}_\text{int}$ describes the local electron-electron interaction within the correlated subspace only, and $\hat{H}_\text{DC}$ is a double counting correction.

In the case of \sco, the correlated subspace for our DMFT calculations is spanned by the partially occupied bands around the Fermi level with dominant Cr-$t_{2g}$ character (see \pref{sec:bandstructure}), expressed in a basis of maximally localized Wannier functions~\cite{Marzari2012} with three $t_{2g}$-like orbitals on each Cr site.
The local interaction $\hat{H}_\text{int}$ is then expressed in the fully rotationally invariant Slater-Kanamori form~\cite{Georges2013}, including spin-flip and pair hopping terms:
\begin{equation}
\begin{aligned}
\hat{H}_{int} &=  U \sum_{m} \hat{n}_{m\uparrow} \hat{n}_{m\downarrow}\\
&+ \sum_{m\neq m'} \Big( (U-2J) \hat{n}_{m \uparrow} \hat{n}_{m' \downarrow} +\sum_{\sigma= \uparrow, \downarrow} (U-3J) \hat{n}_{m \sigma} \hat{n}_{m' \sigma} \Big)  \\
&- \sum_{m\neq m'} J \hat{c}_{m\uparrow}^{\dagger} \hat{c}_{m\downarrow} \hat{c}_{m'\downarrow}^\dagger\hat{c}_{m'\uparrow} +
\sum_{m\neq m'} J \hat{c}_{m\uparrow}^{\dagger} \hat{c}_{m\downarrow}^\dagger \hat{c}_{m'\downarrow} \hat{c}_{m'\uparrow} \quad .
\end{aligned}
\label{eqn:kanamori}
\end{equation}
Here, $U$ and $J$ denote the inter-orbital Coulomb repulsion and Hund's interaction, respectively, while $\hat{c}_{m\sigma}^{\dagger}$ and $\hat{c}_{m\sigma}$ are the creation and annihilation operators for electrons with spin $\sigma$ in orbital $m$, and $\hat{n}_{m\sigma}=\hat{c}_{m\sigma}^{\dagger}\hat{c}_{m\sigma}$ is the corresponding number operator.

The double counting correction, $\hat{H}_\text{DC}$, is supposed to subtract the effect of the local electron-electron interaction within the correlated subspace that is already included in $\hat{H}_\text{KS}$.
However, the explicit form of $\hat{H}_\text{DC}$ is not uniquely defined \cite{Karolak2010}. In this work we follow~\cite{Held2007} where the double counting correction is based on the average interaction energy in the atomic limit.
This results in a constant shift of the self-energy on the Cr sites that can be written as:
\begin{equation}
\Sigma_\text{DC} =  \bar{U}(n-1/2) \quad , 
\label{eqn:DC_sigma}
\end{equation}
where $n$ is the total occupation of the corresponding site and $\bar{U}=U-2J$ is the average interaction strength for the three-orbital Kanamori interaction (considering only density-density terms).

Within DMFT, each correlated site $i$ is mapped on an effective impurity model, thereby neglecting the $k$-dependence of the self-energy \mbox{$\Sigma({\mathbf k}, \omega ) \rightarrow \Sigma(\omega)$} in the local basis~\cite{Georges_et_al:1996}.
The effective impurity model is determined self-consistently by requiring that the corresponding impurity Green's function, $G_\text{imp}(\omega)$, is equal to the local part of the full lattice Green's function, $G({\mathbf k}, \omega)$ for all correlated sites~\cite{Georges_et_al:1996}.

In a fully charge self-consistent DFT+DMFT calculation, the charge density determining the Kohn-Sham potential is recalculated from the local occupations obtained within the DMFT step, and then the procedure is iterated until self-consistency is achieved. However, sometimes, for computational efficiency, such full charge self-consistency is neglected and so-called ``one-shot'' calculations, i.e., without recalculating the charge density and Kohn-Sham potential from the DMFT occupations, are performed. 
In the latter case, there is an ambiguity on whether the double counting correction in \pref{eqn:DC_sigma} is evaluated using the site occupations computed at the DFT level, $n = n_{\text{DFT}}$, or the corresponding DMFT occupations, $n=n_{\text{DMFT}}$~\cite{Hampel2020}. 
While the DMFT occupations should be considered as results of the DFT+DMFT calculations and thus as the actual physically relevant occupations, $n_\text{DFT}$ are the occupations that give rise to the charge density entering the Kohn-Sham potential within a one-shot calculation. In this case it might therefore be more appropriate to use $n_\text{DFT}$ to account for the electron-electron interaction contained at the DFT level.
Note that in charge self-consistent calculations no ambiguity arises, since then $n_\text{DMFT}$ are the physically relevant occupations that also determine the charge density from which the Kohn-Sham potential is calculated.

In the case of of a charge disproportionated system, the double counting correction in \pref{eqn:DC_sigma} directly contributes to the difference of the one-particle energies between sites with different total $d$ orbital occupations. 
Since $\Sigma_\text{DC}$ is subtracted from the Kohn-Sham energies, the double counting correction lowers the energy of the more occupied site relative to the less occupied site, and thus a difference in these site occupations favors the charge disproportionation. 
Therefore, we explore both choices for $n$ in \pref{eqn:DC_sigma} in our initial one-shot calculations presented in \pref{sec:phase_diagram}, while in all our charge self-consistent calculations we always use $n=n_\text{DMFT}$.   

\subsection{\label{sec:Computational_details}Computational Details}

We perform all DFT, \dftu, and DFT+DMFT calculations employing the QuantumESPRESSO package~\cite{Giannozzi2009} (version 7.2) with ultrasoft pseudopotentials (GBRV library~\cite{Garrity2014_GBRV_pseudos}) and the PBEsol exchange correlation functional~\cite{Perdew2009_PBEsol}.   
All results are obtained for the 15 atom hexagonal unit cell described in \pref{sec:structure}, with lattice vectors fixed to a lattice constant of 3.77~\AA\ of the underlying cubic perovskite structure, which corresponds to the relaxed PBEsol value for cubic \sco.
We use a plane wave cutoff of 70~Ry for the wavefunctions and 840~Ry for the charge density. Brillouin zone integrations are performed on $\Gamma$-centered $6 \times 6 \times 6$ Monkhorst-Pack k-point grid and the Marzari-Vanderbilt ``cold-smearing'' scheme~\cite{Marzari1999} with a broadening parameter of 0.001~Ry.

The correlated subspace for the DFT+DMFT calculations is constructed using maximally localized Wannier functions generated with Wannier90~\cite{Pizzi2020}. The DFT+DMFT workflow is implemented using solid\_dmft~\cite{Merkel2022, Beck2022} based on TRIQS/DFTTools~\cite{Parcollet2015, Aichhorn2016}, where the three Cr sites in the cell are mapped on three effective impurity models, which are then solved independently using the continuous time quantum Monte Carlo solver CT-HYB included in TRIQS~\cite{Seth2016}. We use 8000 warmup cycles followed by 10$^7$ Monte-Carlo cycles of 120 moves each at an inverse temperature of $\beta=40$ eV$^{-1}$ (approximately 300 K). The resulting Green's functions are represented using 35 Legendre coefficients~\cite{Boehnke2011}.

All DFT+DMFT calculations are performed for the spin-degenerate paramagnetic case, while spin polarized \dftu calculations are performed for the FM and d-AFM configuration, with the Hubbard correction applied onto orthonormalized atomic projectors. We also perform full cell relaxation within \dftu using a threshold of 10$^{-3}$~Ry/\AA\ for the forces and 10$^{-4}$Ry for the change in energy.

As a final note, we point out that the Hubbard parameters used for the DFT+DMFT and for the \dftu calculations differ, since different projectors as well as different parameterizations of the local interaction are used by the two methods (three \tg frontier Wannier functions with Kanamori interaction in DFT+DMFT compared to five ortho-normalized atomic $d$ orbitals with Slater parameters in \dftu). Therefore, we denote the Hubbard parameter in the DFT+DMFT calculations as $U$, while we use the notation $\mathcal{U}$ for the \dftu calculations. Finally, we do not use an explicit Hund's interaction parameter in our \dftu calculations, since its effect is implicitly taken into account by the effective $\mathcal{U}$ parameter in the simplified form of the $+\mathcal{U}$ correction according to Ref.~\cite{Dudarev1998}. While we performed calculations for several $\mathcal{U}$ values ranging from 2 eV to 4 eV, in \pref{sec:dft+u} we present only the results for the representative value of $\mathcal{U}=3$~eV.

\section{\label{sec:Results} Results}
\subsection{\label{sec:bandstructure} DFT band structure and choice of correlated subspace}

\begin{figure}
   \centering
   \includegraphics[width=\columnwidth]{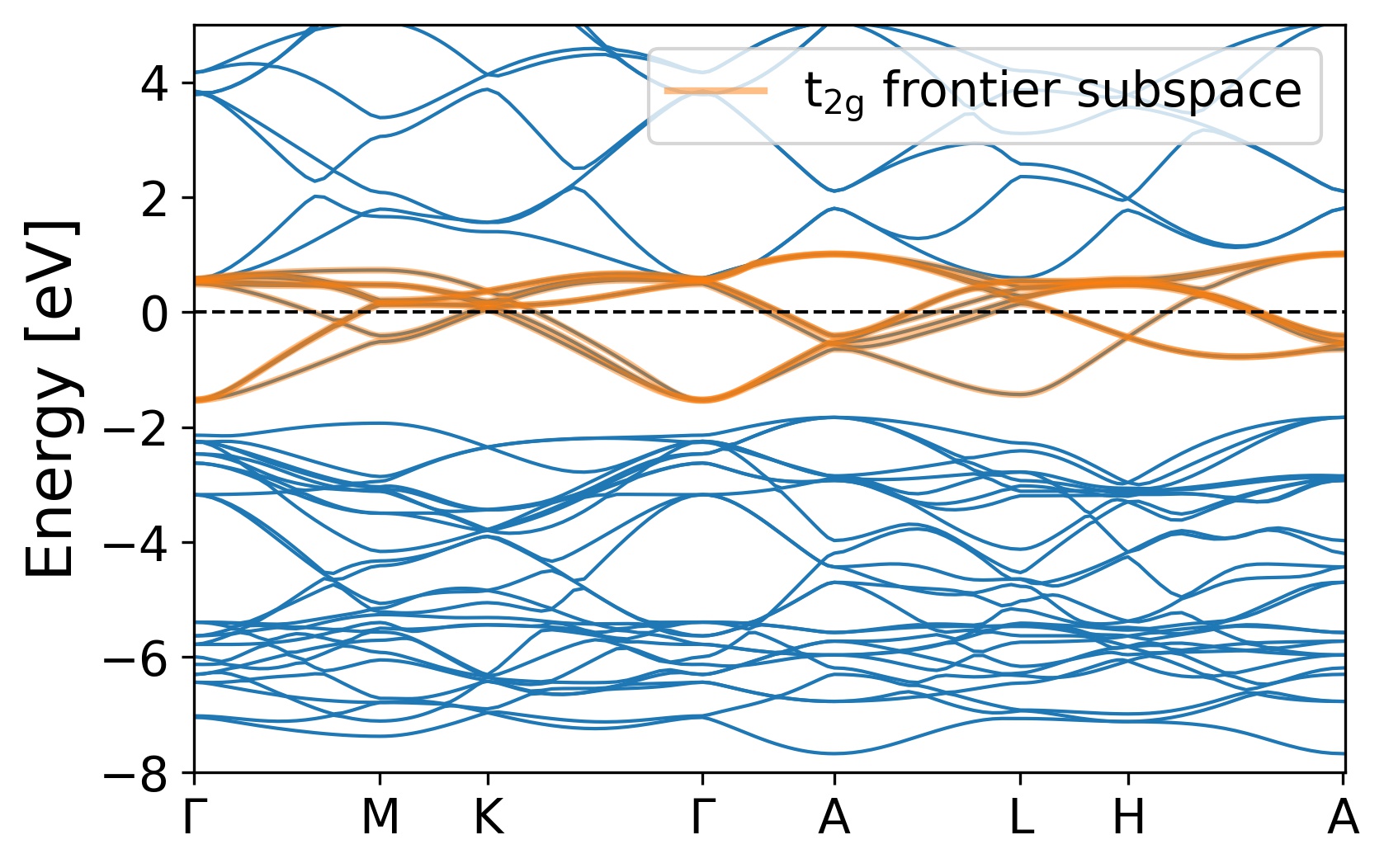}
   \caption{Kohn-Sham band structure of cubic \sco obtained within spin-degenerate DFT (blue lines). The orange lines show the dispersion of the Wannier bands corresponding to the Cr $t_{2g}$ frontier orbitals.}
   \label{fig:bands}
\end{figure}

\pref{fig:bands} shows the spin-degenerate Kohn-Sham band structure obtained for cubic \sco represented in our 15 atom hexagonal supercell. 
The bands around the Fermi level, from slightly above $-2$~eV to about 1~eV, have predominant Cr $t_{2g}$ character, consistent with the formal $d^2$ configuration of the Cr$^{4+}$ cation. 

We construct nine maximally localized Wannier functions from these bands using an energy window between $-2$~eV and 2~eV, starting from initial projections on three $t_{2g}$ orbitals on each Cr site. The resulting Wannier functions still resemble such atomic $t_{2g}$ orbitals centered on the Cr, but also contain $p$-like tails on the surrounding oxygen ligands, representing the $\pi^*$-type anti-bonding character of the hybridization with the O-$p$ orbitals in the underlying bands. The dispersion recalculated from these Wannier functions is highlighted in orange in \pref{fig:bands}.

\subsection{\label{sec:phase_diagram} Phase diagrams for cubic \sco from one-shot DFT+DMFT}

We first describe results obtained within one-shot DFT+DMFT calculations using the DFT occupations to evaluate the double counting correction, i.e., $\Sigma_\text{DC}(n) = \Sigma_\text{DC}(n_{\text{DFT}})$.
Since, in these one-shot calculations, all Cr sites are equivalent at the DFT level, the corresponding occupations are also identical and thus the double counting correction just leads to an overall shift of the chemical potential, with no effect on the result.
These calculations are therefore very similar to the model calculations of Ref.~\cite{Isidori2020}, except that we use the bandstructure of a real material and treat the local interaction within DMFT.

\begin{figure}
   \centering
   \includegraphics[width=\columnwidth]{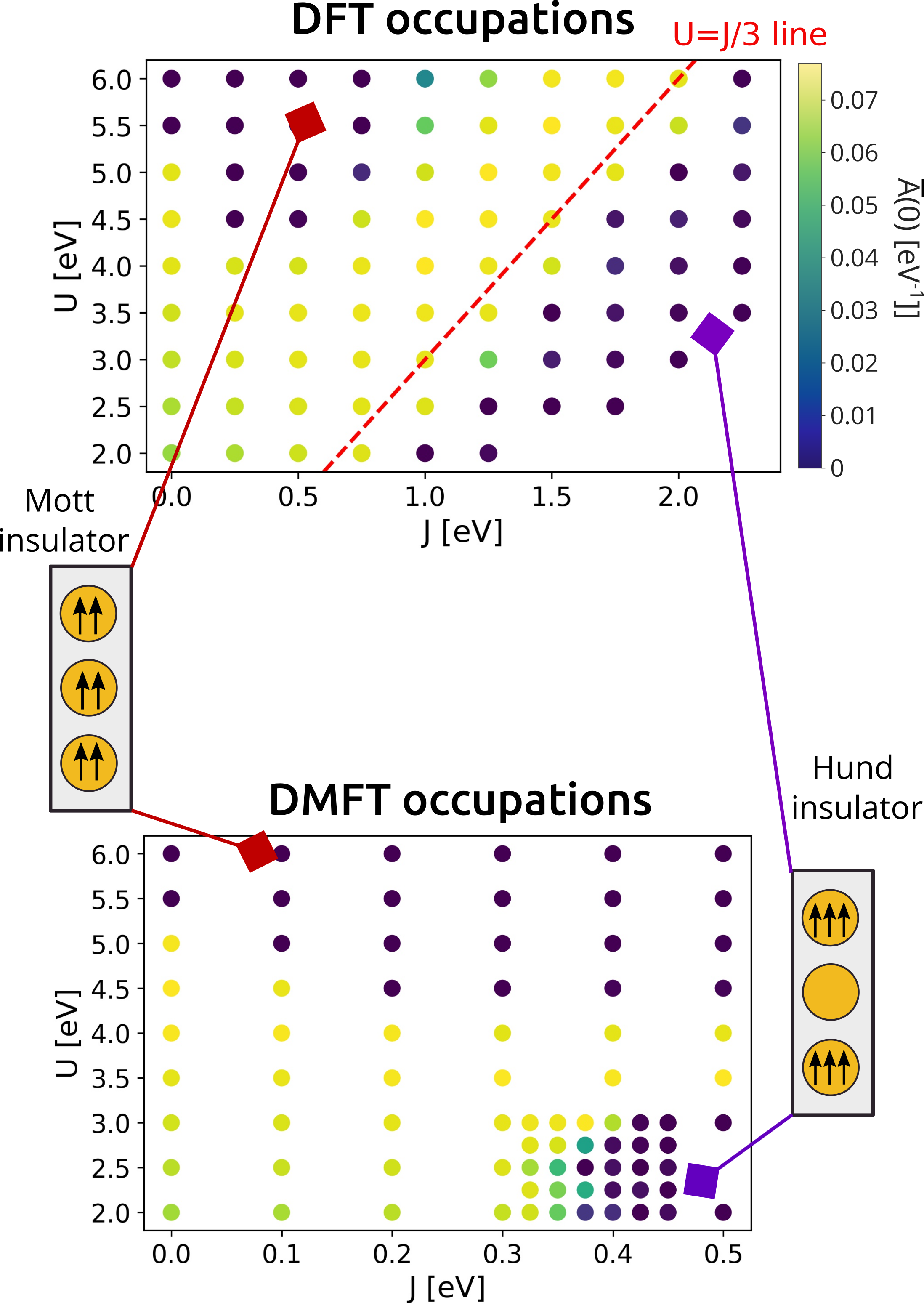}
   \caption{Scatter plot of the value of the spectral function $A(\omega=0)$ evaluated at the Fermi level, for two different choices of the occupations in the double counting correction. Yellow dots correspond to a metallic solutions while dark blue ones denote the presence of a gap. Using DFT occupations \textbf{(a)} in the double counting formula qualitatively reproduces the results obtained for the three-orbital Hubbard model~\cite{Isidori2020}, while employing DMFT occupations \textbf{(b)} shifts the onset of the Hund insulator phase to much lower values of the exchange coupling $J$.}
   \label{fig:phase_diag}
\end{figure}

\pref{fig:phase_diag}(a) shows the value of the averaged spectral weight at zero frequency, $\bar{A}(0) = -\frac{\beta}{\pi}\mathrm{Tr}[G(\beta/2)]$ as function of the interaction parameters $U$ and $J$. 
Non-zero $\bar{A}(0)$ denotes a metallic solution, while a value close to zero indicates an insulating state. 

The phase diagram exhibits the qualitative features that have been reported in the literature for a generic three orbital Hubbard model with an average filling of two electrons per site~\cite{deMedici2011, Georges2013, Isidori2020}. 
A Mott-insulating region, where all Cr sites remain equivalent, is found for large $U$ in the region $U>3J$. The corresponding phase boundary as function of $J$ exhibits the ``Janus effect''~\cite{deMedici2011}, i.e., increasing $J$ from zero first favors the Mott-insulating state, while for $J>0.5$~eV a further increase of $J$ favors the metallic state. In the region $U<3J$ we observe a second insulating phase, where the Cr sites indeed disproportionate according to \dispro, corresponding to the ``Hund's insulator'' reported in Ref.~\cite{Isidori2020}. 
Finally, along the line with $U=3J$, where the Mott and Hund's insulators are energetically degenerate in the atomic limit, the system remains metallic up to high $U$, in agreement with the persistent Hund's metal discussed in \cite{Isidori2020}.

We quantify the degree of disproportionation in the Hund's insulating region by defining the \emph{site polarization} $\Delta n$ as the occupation difference of the more and less occupied Cr sites.
We observe a strong site polarization of $\Delta n > 2.6 $ for essentially all data points within the Hund's insulating region, indicating a relatively abrupt change across the phase boundary to the metallic phase. 

On a technical note, in the absence of the $\Lambda_1$ mode ($Q=0$) and without considering charge self-consistency or using $n_\text{DMFT}$ in the double counting, all three Cr sites remain symmetry equivalent at the single particle level, and thus the specific spatial arrangement of $d^0$ and $d^3$ sites does not provide a significant bias to the input of the effective DMFT impurity problems. This leads to convergence issues in the Hund's insulating regime, where we observe that sites with $d^0$ and $d^3$ filling can arbitrarily switch during DMFT iterations. We do not consider this as problematic, since the calculation always shows insulating behavior in the corresponding regime.

\pref{fig:phase_diag}(b) shows the analogous phase diagram (with a different scale for $J$) obtained by using the DMFT occupations for the double counting correction, i.e., $\Sigma_\text{DC}(n) = \Sigma_\text{DC}(n_\text{DMFT})$. 
While the Mott insulating phase is unaffected by the different double-counting correction, the most striking difference to the previous case is a strong reduction of the minimal $J$ necessary for the onset of the Hund's insulating phase. For $U=2.5$~eV, the disproportionated phase emerges around $J=0.35$~eV. For comparison, using $\Sigma_\text{DC}(n) = \Sigma_\text{DC} (n_{\text{DFT}})$ results in a critical $J$ of 1.25~eV at the same $U$ value.

This massive reduction of the critical $J$ is due to the occupation dependent double counting correction, which, as discussed in \pref{sec:dmft}, results in a lowering of the energy levels on the more occupied relative to the less occupied Cr site, and thus provides a strong positive feedback loop that stabilizes the charge disproportionated phase for much smaller $J$ values than for $\Sigma_\text{DC}(n)=\Sigma_\text{DC}(n_\text{DFT})$. The difference in the double counting correction on the different Cr sites, $\Delta \Sigma_\text{DC} = \bar{U} \Delta n$, reaches up to 5~eV for $U=2.5$~eV and $J=0.4$~eV. 

We note that previous calculations of the interaction parameters for \sco using the constrained random phase approximation (cRPA) obtained $U=2.7$ eV and $J=0.42$~eV~\cite{Vaugier2012}. This would potentially put \sco in the region of the parameter space where the material would spontaneously disproportionate.
However, the results presented so far do not include the effect of charge self-consistency, i.e., the DFT response to the charge correction obtained from DMFT is absent. 
Indeed, when performing such fully charge self-consistent DFT+DMFT calculations for \sco, 
we find that the entire Hund's insulating region in the phase diagram disappears and the metallic region extends over the whole region with $U<4.5$~eV in \pref{fig:phase_diag}(b).

To better understand this strong difference between the one-shot and the charge self-consistent calculations, we now analyze how the charge disproportionation is affected by the structural distortion ($\Lambda_1$ mode) described in \pref{sec:structure}.

\subsection{Effect of the structural distortion}
\label{sec:results-dmft-distortion}

In general, to understand the interplay between electronic and structural degrees of freedom, one may consider a simplified picture where the free energy of the system contains three contributions~\cite{Peil2019, Georgescu2022}:
\begin{equation}
    \mathcal{F}(\Delta n, Q) = \mathcal{F}_{el}[\Delta n] -  \frac{g}{2} Q \Delta n + \frac{K}{2} Q^2 \quad .
    \label{eqn:landau_free_energy}
\end{equation}
Here, $\mathcal{F}_{el}$ depends only on the electronic degree of freedom, which, for the present case,  can be represented in terms of the site polarization $\Delta n$. The second term describes a linear coupling between $\Delta n$ and the structural distortion mode $Q$, while the last term represents the structural mode stiffness. In the case of \sco, $Q$ denotes the normalized amplitude of the $\Lambda_1$ mode shown in \pref{fig:3d_dispro_distor}(b). 

By minimizing the free energy in \pref{eqn:landau_free_energy} with respect to $Q$, one obtains~\cite{Peil2019}:
\begin{equation}
\frac{2K}{g} Q = \Delta n (Q) \quad ,
    \label{eqn:saddle_point}
\end{equation}
where $\Delta n(Q)$ is the site polarization obtained by minimizing the free energy for fixed $Q$.
Thus, if the electronic response to the distortion described by $\Delta n(Q)$ is linear in $Q$, \pref{eqn:saddle_point} has only one solution for $Q=0$ and the structural distortion is unfavorable.
However, if the response of the system is non-linear and, in particular, if there is a strong increase of $\Delta n(Q)$ followed by an approach to saturation, the functions described by the left and right hand sides of \pref{eqn:saddle_point} can intersect at a value $Q \neq 0$ and the system exhibits an energy minimum at non-vanishing distortion.  

\begin{figure}
   \centering
   \includegraphics[width=\columnwidth]{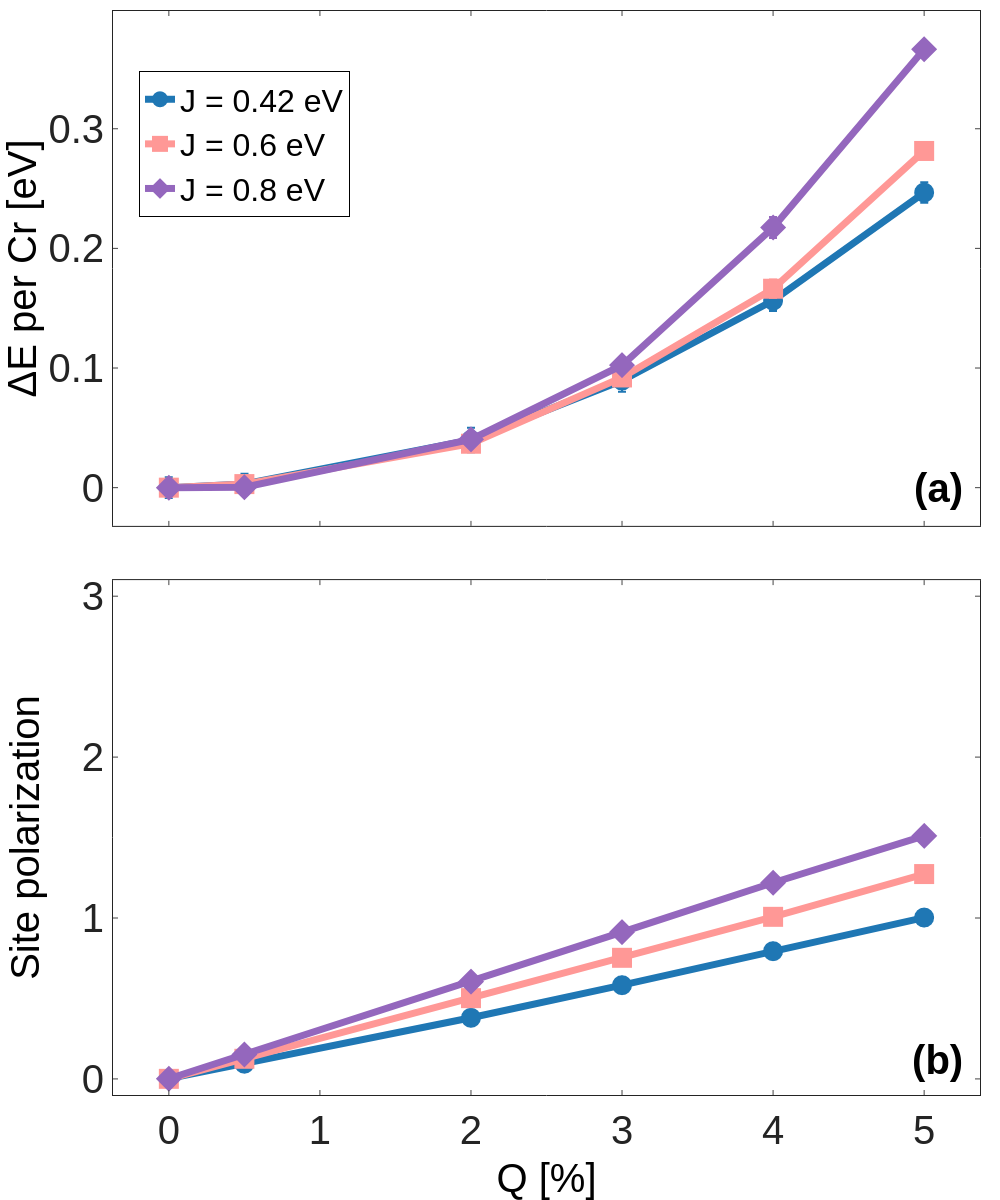}
   \caption{(a) Energy difference per Cr atom between distorted and undistorted \sco and (b) site polarization $\Delta n$ as a function of the $\Lambda_1$ distortion amplitude $Q$ obtained from fully charge self-consistent DFT+DMFT calculations for different $J$ (and $U=2.7$ eV). 
   \label{fig:distorion_J_scan}}
\end{figure}

In \pref{fig:distorion_J_scan} we plot the total energy as well as the site polarization as a function of the normalized mode amplitude $Q$ obtained from fully charge self-consistent DFT+DMFT. 
Data are shown for the cRPA values of the interaction parameters, $U=2.7$~eV and $J=0.42$~eV~\cite{Vaugier2012}, as well as for two higher values for the Hund's coupling $J$.
In all cases, the only minimum of the total energy occurs at $Q=0$, i.e., the structural distortion is energetically unfavorable. 

Simultaneously, the site polarization $\Delta n(Q)$ shows only a linear response to the distortion. Even though this linear response is relatively strong and leads to $\Delta n \approx 1$ for 5~\% distortion, the system always remains metallic so that there is no energy gain related to a gap opening in the spectral function.
This behavior persists even if the Hund's coupling is significantly raised beyond the cRPA value, although $\Delta n$ increases with $J$. 
We note that $J=0.8$~eV is much higher than the unscreened value, $J_\textrm{bare}=0.55$~eV, obtained in~\cite{Vaugier2012}, and thus can be considered as unrealistically high.

\begin{figure}[t]
   \centering
   \includegraphics[width=0.8\columnwidth]{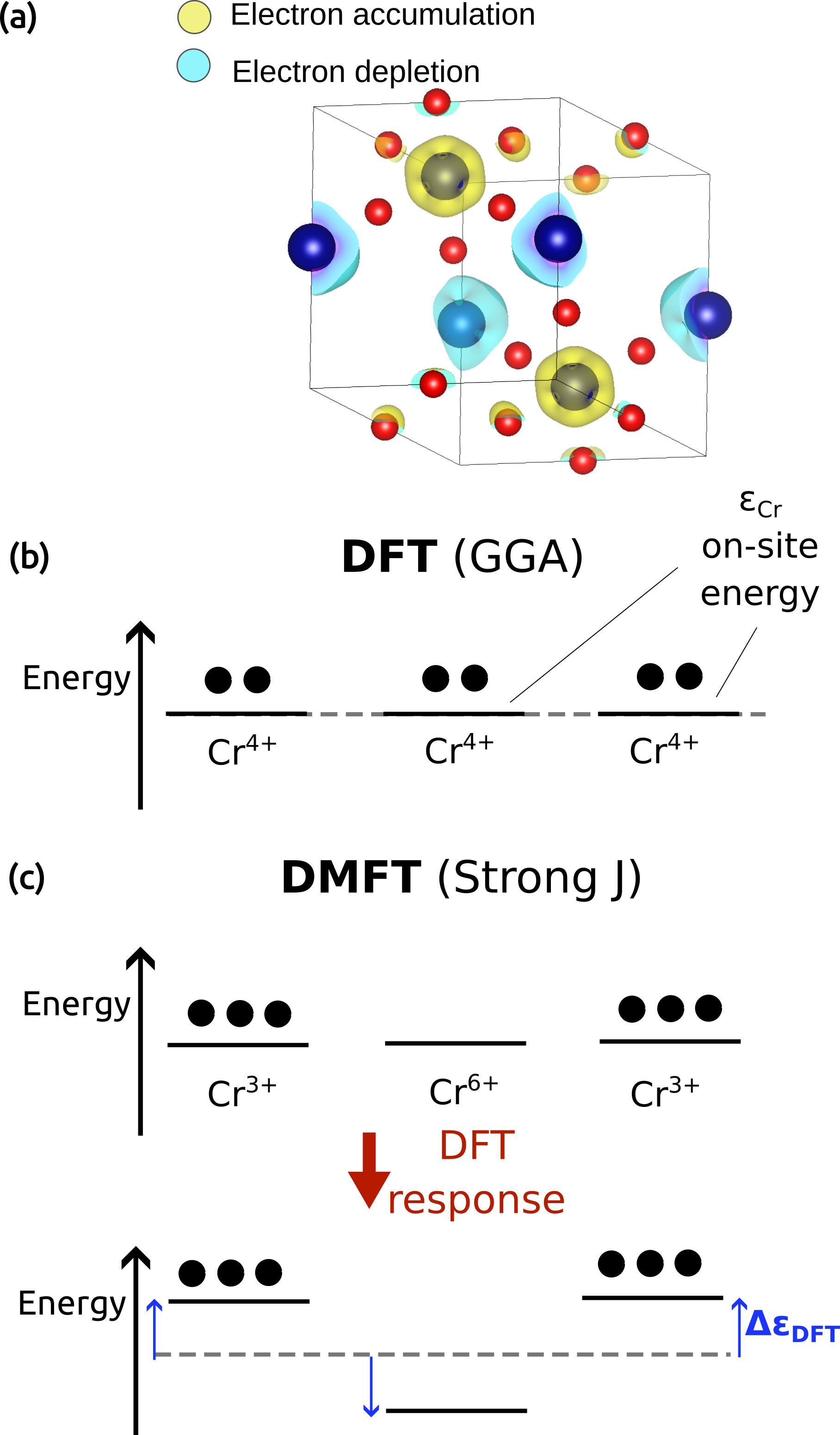}
   \caption{a) DMFT charge correction (isosurface level: 0.00644 $\mathrm{\AA}^{-3}$) after converging the one-shot  calculations for the disproportionated phase \dispro. Dark blue (red) spheres represent the Cr (O) atoms.  (b-c) Schematic picture of the effect of charge self-consistency for the charge disproportionated phase of undistorted \sco. (b) In the initial DFT calculation, all Cr sites are equally occupied. In the undistorted case, all sites are symmetry equivalent so that also the on-site energy is the same. (c) One-shot DMFT converges a charge disproportionated state. This changes the on-site energies within the subsequent DFT step, opposing the tendency towards charge disproportionation.}
   \label{fig:DC_correction_schematic}
\end{figure}
Thus, our results so far indicate no tendency towards charge disproportionation in \sco.

In \pref{fig:DC_correction_schematic}(a), we plot the DMFT correction for the charge density in the disproportionated state obtained from the converged one-shot calculation for $U=2.7$~eV, $J=0.42$~eV, and $Q=0$ (using $\Sigma_\text{DC}(n_\text{DMFT})$). One can clearly see how the \tg orbitals around the nominal Cr $d^0$ sites get depleted while more electrons accumulate mostly on the Cr $d^3$ sites and, to a lower extent, also on the oxygen atoms that are shared between two $d^3$ octahedra.

The regions of accumulated and depleted charge lead to significant changes in the Kohn-Sham potential, which translates to strong shifts of the local atomic-like levels (the on-site term of the one-particle Hamiltonian defined by the Wannier projections). 
In particular, charge accumulation pushes the corresponding local level up in energy, while depletion does the opposite. This is schematically illustrated in \pref{fig:DC_correction_schematic}(b) and (c). 
Thus, the recalculation of the Kohn-Sham potential based in the DMFT-corrected charge density strongly disfavors the charge disproportionation, whereas, as discussed in \pref{sec:phase_diagram}, the use of the more physical DMFT occupations for the double counting correction strongly supports it.
The balance between these two effects therefore determines whether the final state is indeed charge disproportionated. 

One can assume that these DFT level shifts are dominated by the strong local Hartree interaction, even though inter-site Coulomb interactions also contribute.
This local component of the $d$-$d$ interaction is exactly what should be compensated by the double-counting correction. 
However, since the available forms for $\Sigma_\text{DC}$ are based on rather heuristic arguments, these may easily overcompensate or undercompensate the DFT Hartree shift.

In the present case, the double counting correction is constructed from the atomic limit, which might not be very accurate for the relatively delocalized frontier Wannier orbitals used to define the correlated subspace. Furthermore, the use of frontier orbitals leads to essentially nominal $d$ occupations and thus maximizes the resulting site polarization $\Delta n$. While this is attractive for an intuitive physical interpretation, it also amplifies any uncertainties in the double counting correction with respect to the resulting relative local energy levels.

\begin{figure}
   \centering
   \includegraphics[width=\columnwidth]{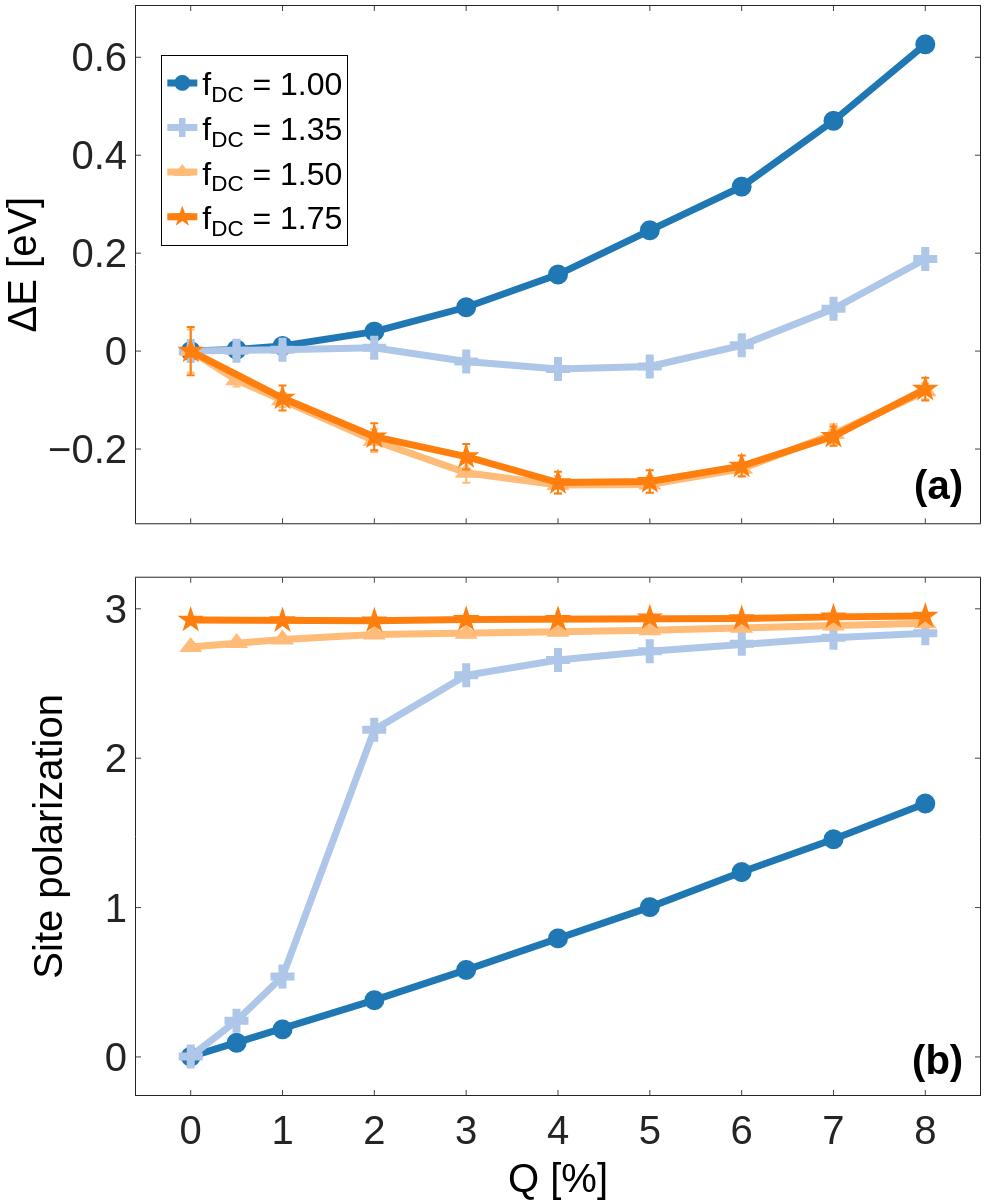}
   \caption{(a) Energy difference per Cr atom between distorted and undistorted \sco and (b) site polarization $\Delta n$ as function of the $\Lambda_1$ distortion mode amplitude for different scaling of the double counting correction ($U$ = 2.7 eV and $J$=0.42 eV). 
   \label{fig:comp_full_DC}}
\end{figure}

We note that there have been many instances reported in the literature, where a scaling of the double counting was necessary to recover the insulating behavior of the studies materials~\cite{Karolak2010, Dang2014, Haule/Birol/Kotliar:2014, Park/Millis/Marianetti:2014}.
Therefore, to investigate how changes in the double counting correction affect the tendency towards charge disproportionation in \sco, we now perform charge self-consistent DFT+DMFT calculations where we scale the double counting correction by a constant factor denoted by $f_{\textrm{DC}}$ while keeping $U$ and $J$ fixed to  2.7~eV a nd 0.42~eV, respectively.
The scaling of the double counting correction can also be viewed as a scaling of the average interaction strength $\bar{U}$ in \pref{eqn:DC_sigma}.
The results are shown in \pref{fig:comp_full_DC}. 

It can be seen that $f_\text{DC} = 1.35$ indeed leads to the formation of an energy minimum at a nonzero distortion amplitude of around 4-5~\%. Simultaneously, the system shows a strong nonlinear increase of the site polarization around $Q=1.5$~\%, which marks the onset of the electronic instability towards formation of the charge disproportionated insulating state ({\it cf.} Ref.~\cite{Peil2019}).
Scaling the double counting correction to 150~\% deepens the energy minimum, and even leads to a spontaneously charge disproprtionated state for zero distortion amplitude with almost saturated $\Delta n$. A further scaling of the double counting correction thus has negligible effect.

These results indicate that \sco is at least close to an instability towards a charge disproportionated insulating phase. However, due to the fundamental uncertainty regarding the size of the double counting correction the predictive capability of the DFT+DFT approach is limited in this case. It is thus not possible to draw a final conclusion whether the charge disproportionated state is indeed a stable state for \sco in ambient conditions.

\subsection{\label{sec:dft+u}  \dftu: structural relaxations and the influence of other distortion modes} 

In the previous sections, we explored the impact of the $\Lambda_1$ distortion mode on the energetics using DFT+DMFT. The observed energy lowering for the scaled double counting correction suggests that the choice of the distortion mode is indeed reasonable. Nevertheless, it is important to also consider other symmetry-allowed distortion modes that might lead to an additional energy lowering. To investigate this, we employ \dftu, which allows us to perform systematic structural relaxations and thus explore a broader range of distortion modes as well as provide further insight into the stability of a charge disproportionated state in \sco.

\begin{figure}
   \centering
   \includegraphics[width=\columnwidth]{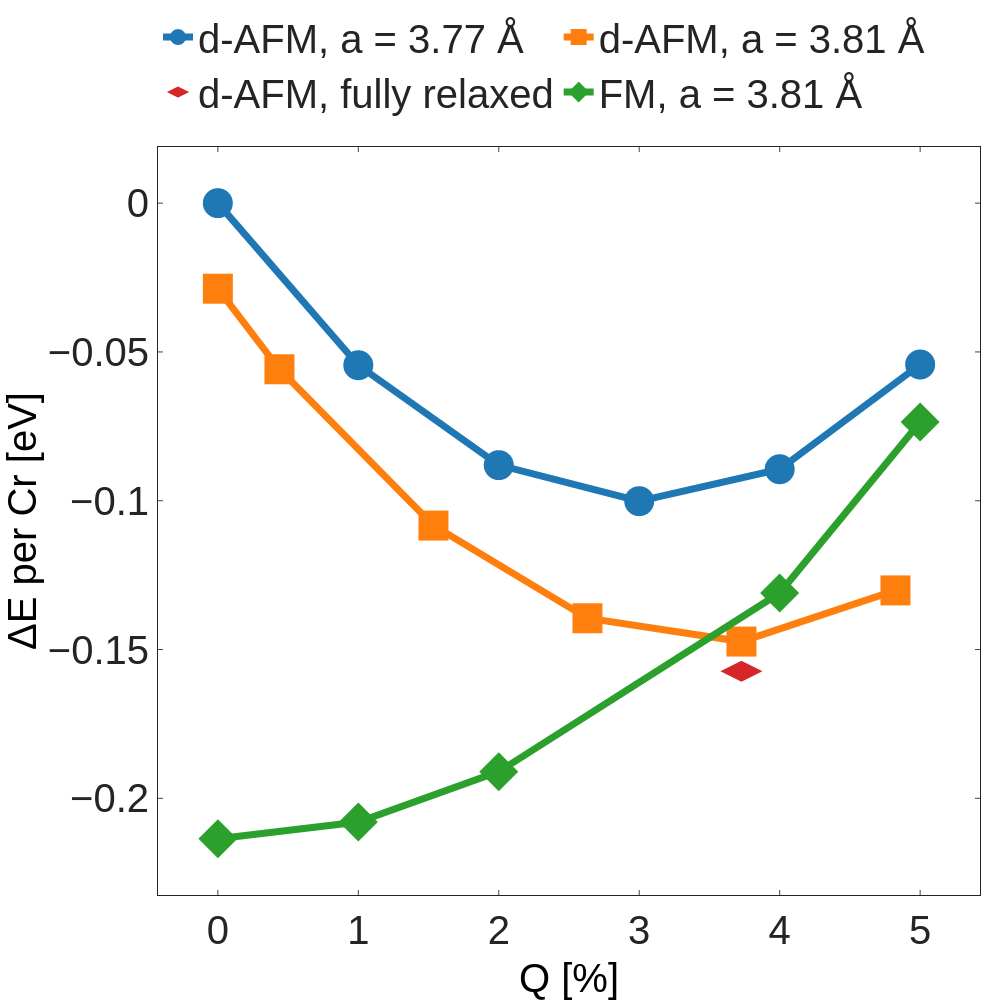}
   \caption{Total energy (per Cr) as function of the $\Lambda_1$ distortion mode amplitude $Q$ obtained from \dftu ($\mathcal{U}=3$ eV) for d-AFM and FM order for different unit cell volumes, indicated by the corresponding pseudocubic lattice constant.  
   The red diamond corresponds to the energy and distortion amplitude obtained from a full structural optimization.
   }
   \label{fig:scan_dist_dft}
\end{figure}

We first perform analogous calculations as for the DFT+DMFT case, where we fix the lattice constant to $a=3.77$~\AA, corresponding to the equilibrium lattice constant of cubic \sco obtained from a spin-degenerate PBEsol relaxation, and then vary the amplitude $Q$ of the $\Lambda_1$ distortion mode. As described in \pref{sec:Methods}, we impose the d-AFM magnetic order in these \dftu calculations to allow for an insulating solution. The resulting energy as function of $Q$ for $\mathcal{U}=3$ eV is shown as blue line  in \pref{fig:scan_dist_dft}.
Indeed, we see that the $\Lambda_1$ mode causes an energy lowering of the system, with a minimum around $Q=3$\%.

\begin{figure}
   \centering
   \includegraphics[width=\columnwidth]{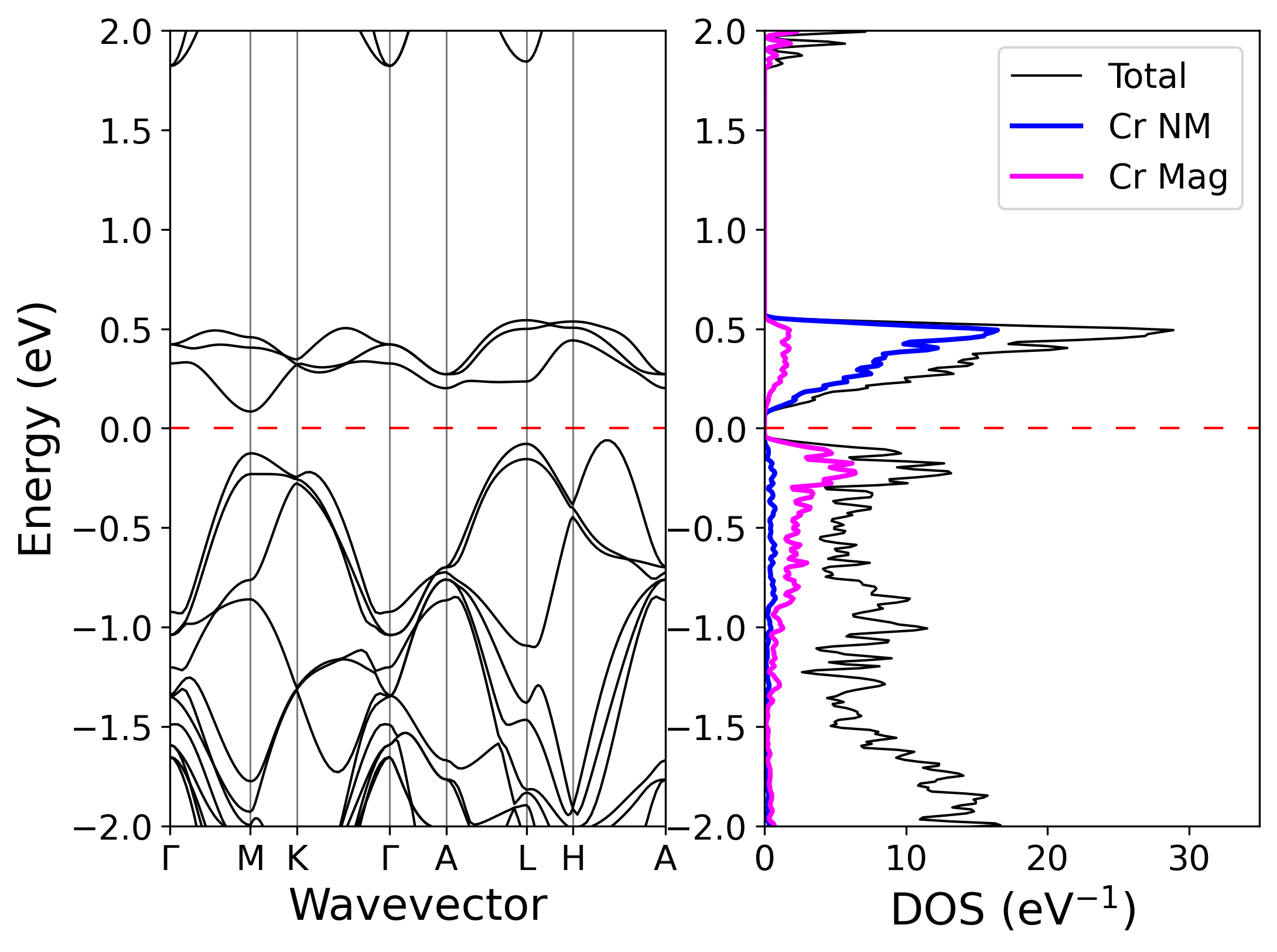}
   \caption{Bandstructure and density of states of \sco (lattice constant $a=3.77$\AA) with d-AFM ($[ \, \cdots \uparrow 0 \downarrow \uparrow 0 \downarrow \cdots \,]$) magnetic order for $\mathcal{U}=3$ eV and $Q$ = 3\%.
   The right panel also shows projections on the two inequivalent Cr sites: the magnetic (Mag, magenta) $d^3$ Cr and the non-magnetic (NM, blue) $d^0$ Cr.}
   \label{fig:dft_bands}
\end{figure}

The d-AFM structure is insulating, as shown in \pref{fig:dft_bands}, where we plot the corresponding density of states (DOS) and band structure for $Q$ = 3\%. 
The projected DOS shows that the top of the valence band region is primarily formed by $d$ states of the nominal $d^3$ Cr, while the unoccupied low-lying conduction band is dominated by states corresponding to the nominal $d^0$ Cr. Thus, even though the difference in the integrated charge on the two inequivalent sites remains relatively small, there is a clear signature consistent with the formal $d^3$ and $d^0$ charge states in the electronic structure. Furthermore, the local magnetic moments are $\sim 2.6~\mu_\text{B}$ on the magnetic $d^3$ Cr and close to zero on the nonmagnetic $d^0$ Cr site, consistent with a high degree of disproportionation between the inequivalent sites.

Next, we perform a full structural optimization (within the $P\bar{3}m1$ space group) for the d-AFM order. The resulting energy and $\Lambda_1$ mode amplitude is marked by the red diamond in \pref{fig:scan_dist_dft}. The relaxation further lowers the energy by about 50~meV and leads to a somewhat larger mode amplitude of $Q \sim 3.8$\%. 
While additional symmetry-allowed distortion modes become non-zero in the relaxed structure, their amplitudes remain negligible compared to the $\Lambda_1$ mode. The main effect leading to the energy lowering is a change in cell volume, while the underlying lattice remains cubic to a good approximation with a pseudocubic lattice constant of $a=3.81$\AA.

To double check the influence of the additional modes, we perform an additional calculation where we fix the lattice vectors to those obtained from the fully relaxed structure, but discard all other distortions except for the $\Lambda_1$ mode. The resulting energy as function of $Q$ is shown by the orange line in \pref{fig:scan_dist_dft},
and exhibits a minimum that is indeed very close to the energy and distortion obtained from the full relaxation. This confirms that the effect of other distortion modes is essentially negligible. 

Finally, we also perform a full structural relaxation for the FM case, and then calculate the corresponding total energy as function of the $\Lambda_1$ distortion amplitude (green line in \pref{fig:scan_dist_dft}). The relaxed cell volume (with $a=3.81$~\AA) for the FM case is essentially identical to that obtained for the d-AFM order. However, the FM calculations always lead to a metallic band-structure and the $\Lambda_1$ distortion is energetically unfavorable in this state. Furthermore, the energy of the undistorted FM state is lower than the energy minimum of the d-AFM state. This indicates that, while a distorted and disproportionated insulating state can be stabilized in these calculations, it is clearly not the energetically preferred ground state of the system.
The d-AFM state also competes with other magnetic phases. For example, the Jahn-Teller distorted C-AFM state explored in Ref.~\cite{Carta2022} is around 0.15 eV per Cr lower in energy compared to the minimum of the FM phase, and 0.2 eV lower when compared to the minimum of the d-AFM case.

\section{Conclusions and outlook}
\label{sec:summary}

To summarize, we have employed both DFT+DMFT and \dftu calculations to explore the possible emergence of a charge disproportionated insulating phase in \sco. We observed that, compared to calculations based on a simple multi-band Hubbard model, the effect of full charge-self consistency in the DFT+DMFT cycle significantly alters the phase diagram obtained as function of the local interaction parameters $U$ and $J$, which can even result in a full suppression of charge disproportionation. Thereby, the stability of the charge disproportionated state is controlled by a subtle balance between the local interaction treated on the DMFT level, which drives the charge disproportionation in particular for $U<3J$, and the DFT response to the DMFT charge correction, leading to a strong Hartree-shift of the local Cr levels that opposes disproportionation.
In practice, this balance is regulated by the double counting correction, for which the explicit form is not clearly defined, therefore limiting the predictive power of the corresponding DFT+DMFT calculations. 

Nevertheless, we have shown that a paramagnetic charge disproportionated insulating state can be stabilized in \sco for realistic interaction parameters via coupling to a structural breathing distortion, if the DFT+DMFT double counting correction is enhanced by about 35~\%.
The breathing distortion and disproportioned state can also be stabilized in \dftu calculations with an appropriate magnetic (d-AFM) order. 
However, comparison with other magnetically ordered states indicates that this d-AFM state is not the ground state configuration of bulk \sco at ambient conditions. Nevertheless, the existence of such a low-energy metastable state, and the existing reports of charge disproportionation in the sister compound PbCrO$_3$~\cite{Wu_et_al:2014, Cheng2015, Zhao_et_al:2023}, suggest that it might be possible to stabilize charge disproportionation also in \sco by appropriate tuning of parameters such as strain, pressure, ionic substitution, or others. 

Our DFT+DMFT calculations also show that a $U$ of at least 4.5~eV is needed for \sco to reach the ``normal'' paramagnetic Mott-insulating region. This is significantly higher than the corresponding values obtained from cRPA calculations~\cite{Vaugier2012}, and suggests that, similar to what has been discussed for BaCrO$_3$ in Ref.~\cite{Giovannetti2014}, a further symmetry breaking, such as, e.g., orbital order or charge disproportionation, is required for \sco to become insulating.

The partially conflicting experimental reports on either metallic or insulating \sco~\cite{Chamberland1967_first_experiment, Komarek2011_experimental_c_type, Long_et_al:2011, Cao2015_paramag_semicond, Bertino2021_MIT_strain}, together with our previous work indicating a C-AFM Jahn-Teller distorted ground state favored by tensile epitaxial strain~\cite{Carta2022}, suggests that \sco is close to many potentially competing structural and electronic instabilities. Further work is thus needed to shed more light on the interplay of magnetism, lattice distortions, and electronic correlations in \sco and other members of this family of perovskite chromates.

\begin{acknowledgments}
This research was supported by ETH Zurich and a grant from the Swiss National Supercomputing Centre (CSCS) under project ID s1128. Calculations were performed on the \enquote{Eiger} cluster of the Swiss Supercomputing Centre (CSCS) and the \enquote{Euler} cluster of ETH Zurich. We thank Maximilian Merkel, Simon J\"ohr and Marta Gibert for valuable discussions.
\end{acknowledgments}

\bibliographystyle{./bibliostyle.bst}
\bibliography{main}
\end{document}